**Quantum entanglement at negative temperature**

*G.B. Furman, V. M. Meerovich, and V.L. Sokolovsky*

Department of Physics, Ben-Gurion University of the Negev, POB 653, Beer-Sheva 84105, Israel

**Abstract**

An isolated spin system that is in internal thermodynamic equilibrium and that has an upper limit to its allowed energy states can possess a negative temperature. We calculate the thermodynamic characteristics and the concurrence in this system over the entire range of positive and negative temperatures. Our calculation was performed for different real structures, which can be used in experiments. It is found that the temperature dependence of the concurrence is substantially asymmetrical similarly other thermodynamic characteristics. At a negative temperature the maximum concurrence and the absolute temperature of the entanglement appearance are sufficiently larger than those at a positive temperature. The concurrence can be characterized by two dimensionless parameters: by the ratio between the Zeeman and dipolar energies and by the ratio of the thermal and dipolar energies. It was shown that for all considered structures the dimensionless temperatures of the transition between entanglement and separability of the first and second spins are independent of spin structure and number of spins.

Quantum entanglement represents one of the most substantial features of the quantum systems and has many prospective applications, for example, in quantum information processing, engineering, communication, metrology [1-3]. However, quantum systems in thermal equilibrium became entangled at very low temperature. This result follows from the fact that quantum systems in thermal equilibrium became entangled if the interaction energy between them $\varepsilon$ is larger than the thermal energy due to their coupling to the environment, [4-7]

$$\frac{\varepsilon}{kT} \propto 1, \qquad (1)$$

where $k$ is the Boltzmann constant, $T$ is the temperature. Typical homonuclear proton-proton dipolar energies in solids are in the range of $\omega_d \propto 10 \div 100\ kHz$. Rough estimation of the temperature of entanglement appearance in a dipolar coupled spin system in equilibrium state gives $T \propto \frac{\omega_d}{k} \propto 0.4 \div 4 \mu K$. Such low temperatures make difficult to use entangled states.

The relationship between temperature and entanglement is not justified for systems that are not in thermal equilibrium [8]. It was shown that, in contrast to the equilibrium cases, entanglement in quantum systems being in a nonequilibrium state can appear at much higher temperatures [8-11]. For example, in a dipolar coupled spin system in a quasi-equilibrium state, when the proton-proton spin system with $\omega_d \propto 100\ kHz$ is described by two temperatures of the Zeeman and dipolar thermal reservoirs, entanglement is achieved at the temperature of the Zeeman reservoir $\propto 23\ \mu K$. For the same system in thermal equilibrium, the temperature of the Zeeman reservoir of the entanglement appearance is $\propto 4 \mu K$ [10].

Determination of the conditions required for the entanglement appearance is important from both practical and theoretical points of view. In the present paper we study entanglement in a quantum 1/2-spin system which thermally isolated from environment and has an upper energy limit of allowed states. Our motivation for taking into consideration of a spin system is based on the fact that many phenomena and systems in quantum physics can be described using spin operator formalism: (i) spin systems have a clear physical picture, (ii) they can be controlled by using a resonance radiofrequency field, (iii) they can be easily measured, for example, nuclear magnetization by the nuclear magnetic resonance (NMR) technique, and (iv) they can be relatively easily described theoretically, since spin operators are defined by simple operation expression and commutation rules.

Let a dipolar coupled 1/2-spin system placed in a constant magnetic field $\vec{H}_0 = H_0 \vec{z}$ is in the equilibrium state. The density matrix of this spin system is

$$\rho = Z^{-1} \exp\left(-\frac{H}{kT}\right) \qquad (2)$$

where $Z = \text{Tr}\{\exp(-H/kT)\}$ is the partition function. The Hamiltonian of the spin system is:

$$H = \omega_0 \sum_{k=1}^{N} I_k^z + H_{dd} \qquad (3)$$

where $\omega_0 = \gamma H_0$ is the Zeeman energy, the energy defined as difference between the excited and ground states of an isolated spin, and $I_k^z$ is the projection of the angular spin momentum operators $\vec{I}_k$ on the z-axes, N is the spin number, $H_{dd}$ describes the spin-spin interactions.

The temperature is defined by the well-known thermodynamic relation [12]

$$\frac{1}{T} = \frac{\partial S}{\partial E} \qquad (4)$$

where

$$S = -kTr(\rho \ln \rho) \qquad (5)$$

is the entropy of the system and

$$E = Tr(\rho H) \qquad (6)$$

is the energy.

The concept of negative temperature was introduced for systems with non-monotonic dependence of the entropy $S$ on the energy $E$: at any point in which the slope of $S(E)$ becomes negative, the temperature $T$ is negative [12-15]. States with negative temperature has been experimentally achieved in a spin system of LiF crystal with dipole-dipole interaction [13].

The concept of negative temperature is physically meaningful for a system that satisfies the following requirements [12-15]: (a) there must be an upper limit to the spectrum of energy states; (b) the system must be in internal thermal equilibrium; (c) the system must be thermally isolated from all systems which do not satisfy both of the conditions (a) and (b).

Here we consider a nuclear spin system in a paramagnetic sample with dipole-dipole interaction described by Hamiltonian $H_{dd}$

$$H_{dd} = -\gamma^2 \hbar \sum_{jk} \left( \frac{3}{r_{jk}^5} (\vec{I}_j \cdot \vec{r}_{jk})(\vec{I}_k \cdot \vec{r}_{jk}) - \frac{1}{r_{jk}^3} \vec{I}_j \cdot \vec{I}_k \right), \qquad (7)$$

where $\gamma$ is the gyromagnetic ratio of the spin, $\hbar$ is the Plank constant, $\vec{r}_{jk}$ is the radius vector from the $j$-th to $k$-th spins with the spherical coordinates $r_{jk}$, $\theta_{jk}$ and $\varphi_{jk}$. $\theta_{jk}$ is the angle between the radius vector $\vec{r}_{jk}$ and the direction of the external magnetic field. The spin system satisfies all conditions required for achieving a negative temperature: (a) the system has a finite number of energy levels, and, hence, the normalization sum converges for any value of $T$ including negative ones; (b) dipole-dipole interactions including the flip-flop terms $I_j^+ I_k^- + I_j^+ I_k^-$ ($I_j^+$ and $I_j^-$ are the raising and lowering spin angular momentum operators of the $j$-th spin) provide establishing thermodynamic equilibrium of the spin system with itself for the time of order of $T_2 \propto \omega_d^{-1} \propto 10^{-5} \div 10^{-4} s$ [16]; and (c) time $T_2$ is much less than the spin-lattice relaxation time $T_1 \sim 100\ s$ and, hence, the spin system can be considered as thermally isolated [16]. At a positive temperature $T > 0$, the equilibrium state is characterized by a greater population of lower energy levels than higher ones. At a negative temperature, the population of the higher levels is greater than that of the lower levels [12-15]. Pictorial schematic diagram of the distribution of spins at different inverse temperatures is presented in Fig. 1.

We will investigate the thermal characteristics and entanglement at negative temperature considering a six- and four-spin circles and six- and eight-spin linear chains which simulate real spin structures. Examples of such systems are dipolar-coupled proton spins of benzene molecule $C_6H_6$ (circle of six spins) [17] and xenon tetrafluoride with chemical formula $XeF_4$ (its crystalline square planar structure was determined by both NMR spectroscopy and by neutron diffraction studies [18, 19]), and calcium hydroxyapatite $Ca_5(OH)(PO_4)_3$ and calcium fluorapatite $Ca_5F(PO_4)_3$ (proton and fluorine chains) [20]. Below we use the dimensionless units: inverse spin temperature $\beta$ =$D/kT$ and the Zeeman energy $\alpha = \frac{\omega_0}{D}$ where $D = \frac{\gamma^2 \hbar}{r_{12}^3}$. The energy $E$ is normalized by the values $\frac{r_{12}^3}{\gamma^2 \hbar}$, and the entropy $S$ and the heat capacity $C$ are normalized by $\frac{1}{k}$. The normalized coupling constants of the $m$-th and $n$-th spins are $\left[\frac{\sin(\pi/N)}{\sin(\pi(m-n)/N)}\right]^3$ for the ring and $1/|m-n|^3$ for the chain, respectively. Now the density matrix and, hence, concurrence are characterized by two dimensionless parameters (scaling parameters): a relation between the Zeeman and dipolar energies $\alpha$ and a dimensionless temperature $\beta$. Numerical calculations are performed using the software based on the MATLAB package. The entropy $S$ and the energy $E$ were calculated using Eqs. (2), (5), (6). Fig. 2 shows the entropy and the inverse temperature as functions of the energy in the dipolar coupling six-spin chain at $\alpha = 1$. The maximum of the entropy separates the areas of the negative and positive temperatures. Due to the spin-spin interaction the plot of $S(E)$ is symmetrical only near $E = 0$, and displays significant asymmetry not so far from $E = 0$. Similar asymmetry is observed in the dependences of the energy, entropy, and heat capacity on the inverse temperature (Fig. 2b). Similar dependences were obtained for all the considered spin systems. For the case of non-interacting spins all the dependencies become symmetrical that agrees with predictions presented in [14].

At $\beta = \infty$ ($T = 0_+$), the system is in its lowest quantum state (Fig. 1, lower line) and its entropy is zero (Fig. 2b). As the inverse temperature decreases, the energy and entropy of the system increase. At $\beta = 0_+$ ($T = \infty$), the entropy reaches a maximum and the specific heat capacity equals to zero. The temperature $T = -\infty$, which corresponds to $\beta = 0_-$, is physically identical to $T = \infty$ ($\beta = 0_+$): these two temperatures give the same spin distribution and the same values of the thermodynamic quantities for the system. A further decrease of $\beta$ corresponds to an increase in temperature from $T = -\infty$ to $T = 0_-$. Finally, at $\beta = -\infty$ ($T = 0_-$), the entropy returns to zero. Thus, the region of negative temperatures lies not below zero but above infinity [12-15]. When a system at a negative temperature is brought into contact with a system at positive temperature, energy will be transfused from the system at negative temperature to the system at positive temperature. It can be said that negative temperatures are "hotter" than positive temperatures [12].

If a nuclear magnetic resonance experiment is carried out on a spin system at negative temperature, the resonant emission of energy is taken place instead of the resonant absorption. To achieve a negative temperature in a spin system placed in an external magnetic field, it is possible to reverse quickly the direction of the field and then to perform adiabatic demagnetization [12, 13].

We will characterize the entangled states by the concurrence between two, the first and second, spins, which is defined as [21]

$$C_{12}(\beta) = \max\{q(\beta), 0\}, \qquad (8)$$

with $q(\beta) = \lambda^{(1)}(\beta) - \lambda^{(2)}(\beta) - \lambda^{(3)}(\beta) - \lambda^{(4)}(\beta)$. Here $\lambda^{(k)}(\beta)$ $(k=1,2,3,4)$ are the square roots of eigenvalues, in the descending order, of the following non-Hermitian matrix:

$$R_{12}(\beta) = \rho_{12}(\beta)(\sigma_y \otimes \sigma_y)\tilde{\rho}_{12}(\beta)(\sigma_y \otimes \sigma_y), \qquad (9)$$

where $\rho_{12}(\beta)$ is the reduced density matrix. For the first and second spins, the reduced density matrix $\rho_{12}(\beta)$ is defined as $\rho_{12} = Tr_{12}(\rho)$ where $Tr_{12}(...)$ denotes the trace over the degrees of freedom for all spins except the first and second spins. In Eq. (9) $\tilde{\rho}_{12}$ is the complex conjugation of the reduced density matrix $\rho_{12}$ and $\sigma_y$ is the Pauli matrix $\sigma_y = \begin{pmatrix} 0 & -i \\ i & 0 \end{pmatrix}$. For maximally entangled states, the concurrence is $C_{12} = 1$, while for separable states $C_{12} = 0$.

Similar to the asymmetry in the temperature dependences of the thermodynamic characteristics (Fig. 2), the temperature dependence of the concurrence is also asymmetrical: at negative temperature the concurrence reaches the value larger than that at positive temperature (Fig. 3). The temperature of entanglement appearance is independent of the structure and the number of spins in the sample (Fig. 3) and is also asymmetrical. At $\beta < 0$ this temperature slowly depends on parameter $\alpha$ and changes by about 10% at increase of $\alpha$ from 0.5 to 1.5 while at $\beta > 0$ the temperature of entanglement appearance varies twice therein this range of $\alpha$.

The spin number and spin structure type as well as parameters $\alpha$ and $\beta$ influence the concurrence magnitude.

To estimate the critical temperature, below which entanglement exists, let us consider fluorine with $\gamma$ = 4.0025 kHz/G in the local field ~ 8 G [16, 22], it corresponds to $T \propto 0.77\,\mu K$ and $T \propto -1.6\,\mu K$. To reach the temperature $T = 0.77\,\mu K$, the sample at the temperature ~ 0.6 mK is magnetized in the field $H_0$ = 6376 G as in the experiment [13] and then the direction of the field should be is reversed so quickly that the spins "cannot follow it" [12, 13]. The system is thus in a non-equilibrium state. During a time of the order of $T_2$, the system reaches an equilibrium state. If the field is then adiabatically reduced to low value of ~ 8 G, the system remains in the equilibrium state [22], which will have a negative temperature. In the final state achieved at adiabatic demagnetization, the parameters were calculated by the way described in detail in [7]. The exchange of energy between the spin system and the lattice, whereby their temperatures are equalized, takes place in a time of the order of $T_1 >> T_2$ [12, 16]. An alternative way to reach the temperature $T = -1.6\,\mu K$ can be implemented as follows: we can take the sample at the temperature of 1.3 mK in the same field, $H_0$ = 6376 G, to magnetize it, and then to reverse the nuclear magnetization using a radiofrequency π-pulse. Finally, the field has to be adiabatically reduced to the same low value ~ 8 G.

In conclusion, we have demonstrated the appearance of entangled states in quantum systems at negative temperatures. The temperature dependence of the thermodynamic characteristics and concurrence are substantially asymmetrical. An unexpected behavior of the nearest-neighbor concurrence was obtained: for all the considered geometric and various number of spins the temperature of the entanglement appearance is independent of the structure (Fig. 3), for both positive and negative temperatures. This temperature is determined by two dimensionless parameters: the ratio $\alpha$ between the Zeeman and dipolar energies and a dimensionless temperature $\beta$ which is the ratio of the dipolar and thermal energy. The concurrence value is determined by along with these parameters also by the system structure and number of spins. At negative temperature the concurrence and temperature (absolute value) of the entanglement appearance are larger than those at positive temperature.

The authors thank the referees for their interest in this work, their comments and suggestions that motivated the authors to perform additional studies with various numbers of spins, various geometry and various ratios between the energies. This study led to obtaining new interesting results.

References

1. G. Benenti, G. Casati, and G. Strini, *Principles of Quantum Computation and Information*, Vols. I and II (World Scientific, Singapore, 2007).
2. L. Amico, R. Fazio, A. Osterloh, and V. Vedral, Rev. Mod. Phys. **80**, 517 (2008).
3. R. Horodecki, P. Horodecki, M. Horodecki, and K. Horodecki, Rev. Mod. Phys. **81**, 885 (2009).
4. S. I. Doronin, E. B. Fel'dman, M. M. Kucherov, and A. N. Pyrkov, J. Phys.: Condens. Matter **21**, 025601 (2009).
5. G. B. Furman, V. M. Meerovich, and V. L. Sokolovsky, Quant. Inf. Proc. **10**, 307 (2011).
6. G. B. Furman, V. M. Meerovich, and V. L. Sokolovsky, Quant. Inf. Proc. **11**, 1603 (2012).
7. G. B. Furman, V. M. Meerovich, and V. L. Sokolovsky, Phys. Lett. A **376**, 925 (2012)
8. F. Galve, L. A. Pachon, and D. Zueco, Phys. Rev. Lett. **105**, 180501 (2010).
9. V. Vedral, Hot entanglement, Nature (London) **468**, 769 (2010).
10. G. B. Furman, V. M. Meerovich, and V. L. Sokolovsky, Phys. Rev. A, **86**, 032336 (2012).

11. G. B. Furman, V. M. Meerovich, and V. L. Sokolovsky, Phys. Rev. A, **78**, 042301 (2008).
12. L.D. Landau and E.M. Lifshitz, *Statistical Physics*, part 1, Third Revised and Enlarged Edition, Pergamon press Oxford, 1980.
13. E. M. Purcell and R. V. Pound, Phys. Rev. 81, 279 (1951).
14. N. F. Ramsey, Phys. Rev. 103, 20 (1956).
15. M. J. Klein, Phys. Rev., 104, 589 (1956).
16. A. Abragam, *The Principle of Magnetic Resonance*, Oxford, Clarendon Press, 1961.
17. J.-S. Lee, T. Adams, and A.K. Khitrin, J. Chem. Phys. **122**, 041101 (2005)
18. T. H. Brown, E. B. Whipple, and P. H. Verdier, Science **140**, 1208 (1963).
19. John H. Burns, P. A. Agron, and Henri A. Levy, Science **139**, 1208 (1963).
20. G. Cho, J. P. Yesinowski, J. Phys. Chem. **100**, 15716 (1996).
21. W. K. Wootters, Phys. Rev. Lett. **80**, 2245 (1998).
22. A. Abragam and W.G. Proctor, Phys. Rev. **109**, 1441(1958).

Figures

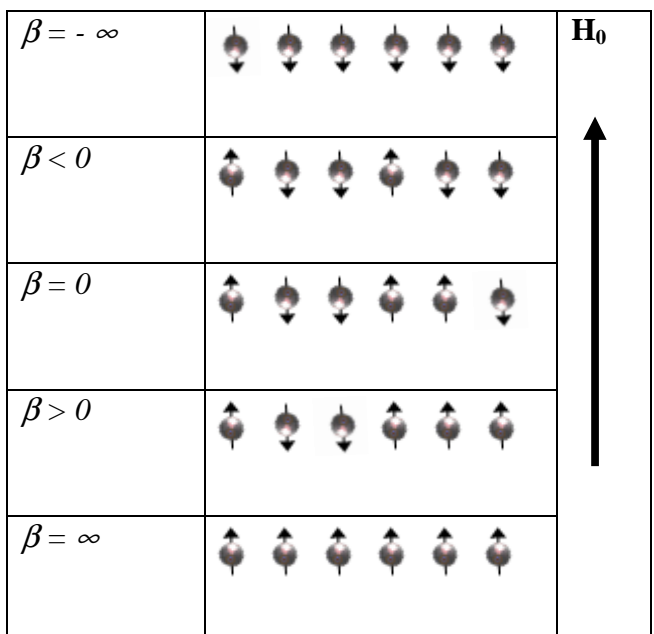

Fig. 1 Spin distribution for positive and negative temperatures. The magnetic field is directed upward.

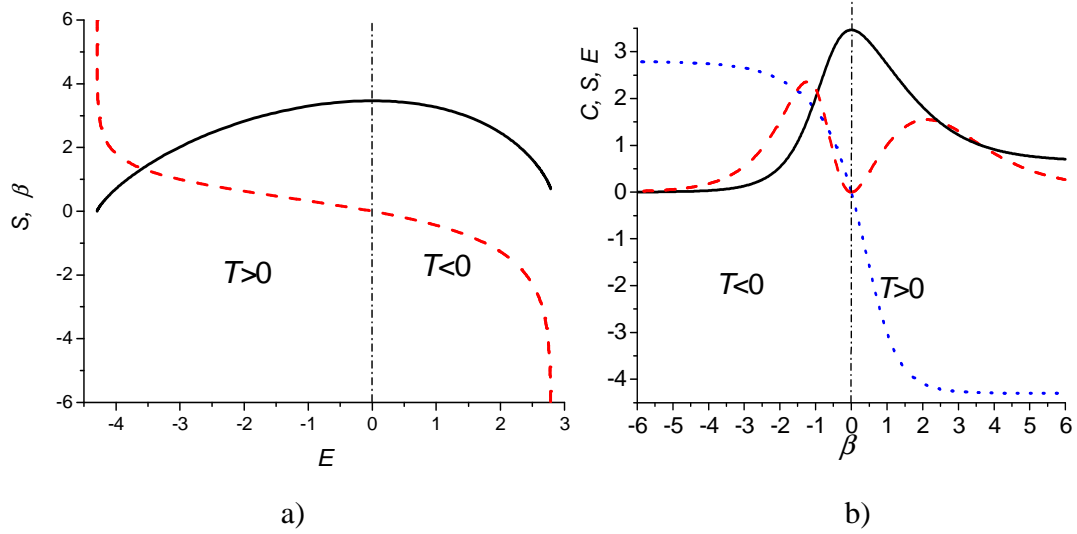

Fig. 2. (Color online) (a) Entropy (black solid line) and inverse temperature (red dashed line) as functions of energy in the 6-spin chain. (b) Entropy (black solid line), energy (blue dotted curve), and heat capacity (red dashed curve) as functions of inverse temperature at $\alpha = 1$. The vertical dash-dotted line separates the areas of positive and negative temperatures.

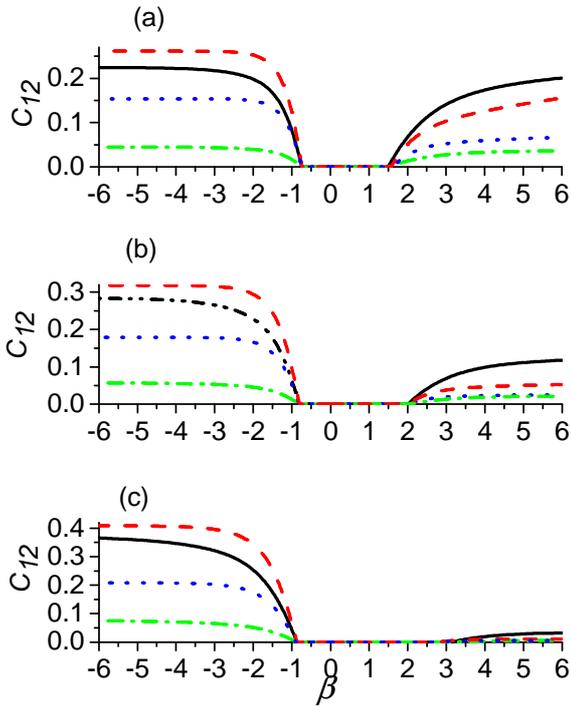

Fig. 3 (Color online) The concurrence between nearest spins as a function of the inverse temperature $\beta$: black solid line – 6-spin chain; dashed red – 4-spin circle; doted blue – 6-spin circle; dash-doted green – 8-spin chain. (a) $\alpha=1.5$, (b) $\alpha=1$, (c) $\alpha=0.5$.